\documentclass{elsart}

\usepackage{harvard}

\usepackage{graphicx}

\usepackage{amssymb}

\def\url#1{{\ttfamily\def\/{/\discretionary{}{}{}}#1}}

\begin{document}

\begin{frontmatter}
\title{The importance of radio sources in accounting for the highest mass 
black holes}


\author[Lacy]{M.\ Lacy\thanksref{ml}}, 
\author[Ridgway]{S.\ Ridgway\thanksref{sr}}
\author[Trentham]{N.\ Trentham\thanksref{nt}}

\thanks[ml]{Present address: IGPP,
Lawrence Livermore National Laboratory,
Livermore, CA 94550.
E-mail: m.lacy1@physics.oxford.ac.uk}
\thanks[sr]{E-mail: ridgway@pha.jhu.edu}
\thanks[nt]{E-mail: trentham@ast.cam.ac.uk}

\address[Lacy]{NAPL, Keble Road, Oxford, OX1 3RH}
\address[Ridgway]{Bloomberg Center for Physics and Astronomy, 
Johns Hopkins University, Baltimore, MD 21218}
\address[Trentham]{Institute of Astromony, Madingley Road, Cambridge, CB3 OHA}

\begin{abstract}
The most massive black holes lie in the most massive elliptical galaxies, 
and at low-$z$ all radio-loud AGNs lie in giant ellipticals. This strongly 
suggests a link between radio-loudness and black hole mass. We argue that 
the increase in the radio-loud fraction with AGN luminosity in 
optically-selected quasar samples is consistent with this picture. We also use 
the ratio of black holes today to quasars at $z\sim 2$ to conclude that 
the most bolometrically-luminous AGN, either radio-loud or radio 
quiet, are constrained to have lifetimes $\stackrel{<}{_{\sim}}10^8$yr.
If radio sources are associated with black
holes of $\stackrel{>}{_{\sim}} 10^9 M_{\odot}$ 
at all redshifts, then the same lifetime constraint applies to 
all radio sources with luminosities above 
$L_{5 {\rm GHz}}\sim 10^{24} {\rm WHz^{-1}sr^{-1}}$.  
\end{abstract}

\end{frontmatter}

\section{Introduction}
\label{intro}

Only $\sim 10$ \% of quasars in optically-selected
samples are radio loud. This fraction does, however, seem to be a 
function of quasar luminosity. Goldschmidt et al.\ (1999) show that
the radio-loud fraction (where radio-loud is defined as having a 
5GHz radio luminosity $L_{5 {\rm GHz}}> 
10^{24} {\rm WHz^{-1}sr^{-1}}$) increases
with quasar luminosity from about 10\% at $M_{\rm B}=-26$ to around
40 \% at $M_{\rm B}=-28$ (Fig.\ 1). As can be seen from the figure, this 
does not seem to be due simply to the radio-quiet quasars becoming 
more radio-luminous with increasing AGN luminosity and crossing the 
threshold into radio loudness (as even radio-quiet 
quasars are not radio silent), but to a genuine 
increase in the relative numbers 
of the powerful ($L_{5{\rm GHz}}>10^{25} {\rm WHz^{-1}sr^{-1}}$) radio sources.

In this paper, we combine this observation with recent studies 
showing that the masses of the spheroidal components of 
galaxies and the mass of the central black hole are correlated
(Magorrian et al.\ 1998; van der Marel 1999). Since at low-$z$
all radio-loud AGNs lie in giant ellipticals, this is strongly suggestive
of a connection between radio-loudness and black hole mass. Thus we can 
now begin to speculate on the importance of radio-loud objects in the evolution
of the most massive black holes.

\begin{figure}
\centering
\includegraphics[scale=0.6,angle=-90]{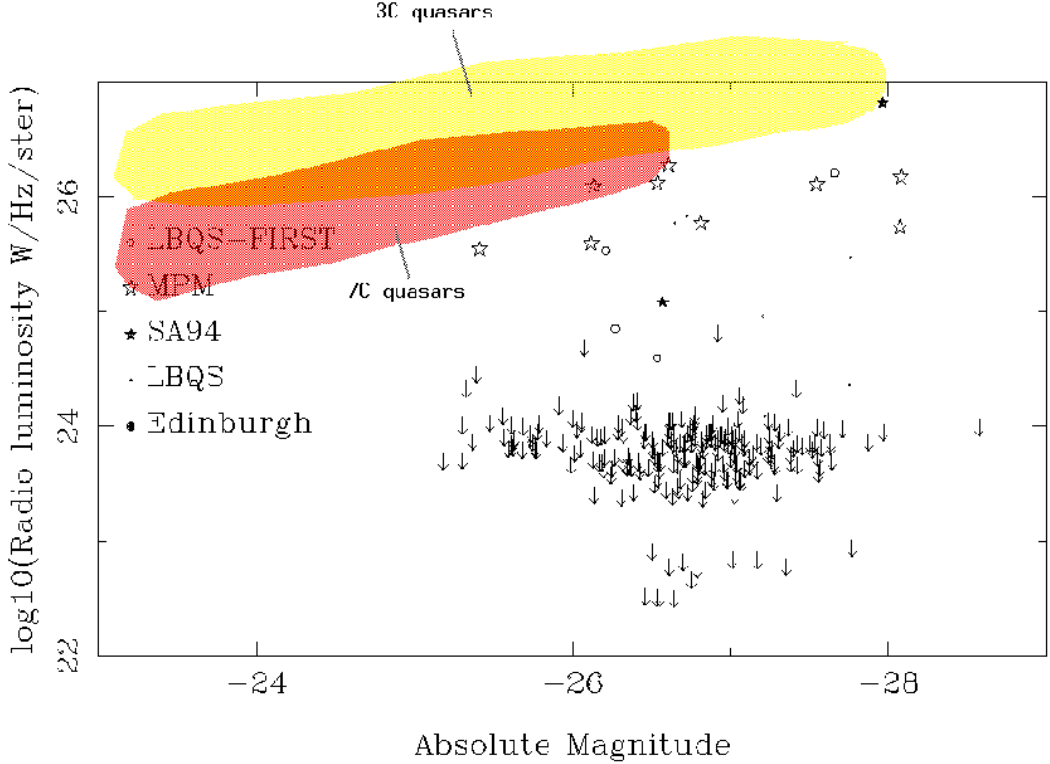}
\caption{Figure 5 of Goldschmidt et al.\ 1999, modified to indicate the 
areas occupied by the radio-selected 3C and 7C quasar samples 
(Willott et al.\ 1998). The 
radio-selected samples naturally favour radio-brighter objects at 
a given optical luminosity, explaining their offset towards the top of the 
plot.}
\end{figure}

\section{Assumptions}

As is usual, we characterise the accretion 
process in quasars in terms of two parameters, the accretion efficiency 
$\epsilon$, usually assumed to be $\sim 0.1$, 
and the ratio of the accretion rate to that at the Eddington limit, 
$\lambda$. We assume a bolometric correction to B-band of 12 
(Elvis et al. 1994) giving a B-band quasar absolute magnitude for a black 
hole of mass $M_{\rm h}$ of 
\[M_{\rm B, Q} = -26.2 + 2.5{\rm lg} 
\lambda - 2.5{\rm lg}(M_{\rm h}/10^9 M_{\odot}).\]

The spheroid luminosity, using van der Marel (1999) relation converted into 
$B$-band magnitudes assuming $B-V=0.9$, is: 
\[M_{\rm B, Host} = 0.555 -2.5{\rm lg} (M_{\rm h}/M_{\odot}).\]
We take $H_0=50 {\rm kms^{-1}Mpc^{-1}}$ and $\Omega=1$ throughout.

\section{Comparing the numbers of black holes to quasars}

By ratioing the number density of black holes of a given mass 
(from Salucci et al.\ 1999) with the 
number density of quasars at the peak of AGN activity at $z=2$ and 
assuming Eddington-limited accretion, one can estimate 
the quasar duty cycle (Fig.\ 2). Given that the quasar epoch lasts 
for $\approx 10^9 {\rm yr}$, this can be converted to a quasar 
lifetime assuming a single burst of activity (Richstone et al.\ 1998).

The uncertainties in the comparison are large, however, and the 
ratios should only be considered order of magnitude estimates. The comparison 
depends on the ratio of the high luminosity/mass ends of two very steeply 
declining functions. Consequently the ratio of number densities is very 
sensitive to errors in the calibration of the black hole mass to bulge and 
quasar luminosity relations. Also any intrinsic scatter in the black hole 
mass -- bulge mass relation will tend to raise the black hole to quasar
ratio when integrating the mass function from a lower bound in black hole 
mass. Finally, the number density of any heavily obscured
radio-quiet ``quasar-2'' population, which is probably required to explain the 
hard X-ray background, is uncertain. It may be about the same as the normal
quasar population (Salucci et al.\ 1999). If so it would be 
comparable to the ratio of 
luminous radio galaxies (i.e.\ the radio-loud quasar-2s) to radio-loud 
quasars and the ratio of type 1 to type 2  Seyferts. However, 
Fabian (1999) suggests that quasar-2s could be 
ten times more common than quasar-1s. 

\begin{figure}
\centering
\includegraphics[scale=0.5,angle=-90]{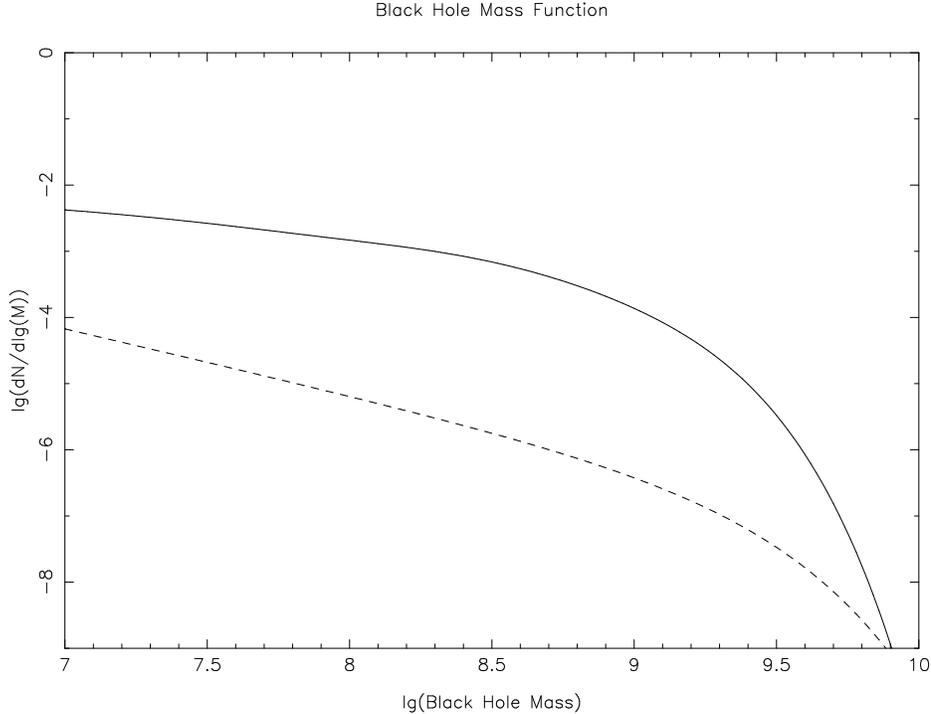}
\caption{A comparison of the local black hole mass function (solid line) 
with the quasar luminosity function at $z=2$ converted into a black hole
mass function assuming Eddington-limited accretion
[dashed line; from Goldschmidt
\& Miller 1998, slightly modified at the faint end to give number densities 
consistent with Boyle (1992)].}
\end{figure}

\section{The most luminous objects}

Goldschmidt et al.\ (1999) show that the number densities
of radio-loud and radio-quiet quasars become comparable at 
$M_{\rm B, Q} \approx -28$. This corresponds to a
black hole of mass $\approx 5\times 10^9 M_{\odot}$ for $\lambda=1$, 
or a host luminosity
today of $M_{B} \approx -23.7$, at the top end of the observed galaxy 
luminosity function and consistent with the most optically-luminous FRI 
hosts seen today. The number density of these black holes is
$\sim 10^{-6}$ Mpc$^{-3}$. 

Using the quasar luminosity function of Goldschmidt \& Miller (1998) we obtain 
a number density of these objects at $z=2$ of 
$\approx 2\times 10^{-8}$ Mpc$^{-3}$ (comoving), 
a factor of $\approx 50$ lower than 
the number density of corresponding black holes now. Adding the
obscured population will raise the number density by a 
factor of 2-10, resulting in a total of $\sim 10^{-7} {\rm Mpc^{-3}}$.
This is an order of magnitude less than the number density of
corresponding black holes in the local Universe, but is a smaller
ratio than for the black holes powering quasars close to the break in the 
black hole mass function where the ratio is around 100 (Fig.\ 2), and perhaps
as high as 1000 (Richstone et al.\ 1998). 

The similarity in number densities of radio-loud and radio-quiet objects 
at these luminosities (unless we assume a very large population of 
radio-quiet quasar-2s) allows us to constrain the lifetimes of the radio-loud
objects in a similar manner to those of the radio-quiets. If they had very 
disparate lifetimes one class or the other would dominate the mass accretion
and produce objects with giant black holes, which are not seen (so far) 
whilst the other class
would be unable to accrete a significant fraction of their black-hole mass.

Thus we conclude that all the most luminous quasars, regardless of 
type, have duty cycles $\sim 0.1$, 
meaning a maximum active lifetime of $\sim 10^8$yr. This agrees with radio 
source ages based on lobe expansion speed estimates
(Scheuer 1995), and allows a moderate amount of growth
of the black hole as a black hole increases in 
mass by a factor $e$ every $t_{\rm S} = 4\times 10^7 (\epsilon/0.1) \, 
{\rm yr}$. 

\section{Lower luminosity objects}

As the quasar luminosity decreases, the radio fraction drops off rapidly, 
to only $\sim$10\% at $M_{\rm B}>-26$. Why? At low redshifts
radio galaxies are seen in giant elliptical hosts, but low luminosity 
radio-quiet quasars can be found in either ellipticals or spirals,
e.g.\ McLure et al.\ 1998. Perhaps
radio-loud objects are found exclusively associated with massive (and 
therefore rare) $>10^9 M_{\odot}$ black holes, accreting at sub-Eddington 
rates whereas low-luminosity radio-quiets continue to be associated with 
Eddington or near-Eddington accretion onto less massive (and 
therefore much more common) black holes.
This picture is supported by stored energy arguments 
(Rawlings \& Saunders 1991), which suggest 
$M_{\rm h}\stackrel{>}{_{\sim}} 10^8 M_{\odot}$
in even low luminosity FRII radio sources.

The number density of $L_{5{\rm GHz}}>10^{24} {\rm WHz^{-1}sr^{-1}}$ 
radio sources is $\approx 5\times 10^{-6}$, 
$\approx 20$ times lower than 
the number density of $>10^9 M_{\odot}$ black holes. Thus 
radio sources with $L_{5{\rm GHz}} \sim 10^{24} {\rm WHz^{-1}sr^{-1}}$
accreting with $\lambda \sim 10^{-3}$ would have 
duty cycles again constrained to be about 10\%, implying lifetimes 
$\sim 10^8$ yr. An interesting alternative possibility is that these 
low luminosity sources are the 
90 \% of the very massive ($> 3\times 10^9 M_{\odot}$) black holes which
have $\lambda \ll 1$ at $z\sim 2$. This would allow them to have duty cycles
$\sim 1$ and lifetimes $\sim 10^9$ yr, which would, however, be rather 
longer than other radio source lifetime estimates.


\end{document}